\documentstyle[epsf]{l-aa}

\begin{document}

\thesaurus{07(08.09.2 HR Car; 08.09.2 $\eta$\,Car; 08.05.3; 08.13.2; 09.02.1; 
09.10.1)}

\title{The bipolar structure of the LBV nebula around HR Carinae}

\author {K.\ Weis \inst{1,2}\thanks{Visiting Astronomer, Cerro Tololo 
Inter-American Observatory, National Optical Astronomy Observatories, 
operated by the Association of Universities for Research in Astronomy, Inc., 
under contract with the National Science Foundation.} 
\and W.J.\ Duschl \inst{1,3} 
\and D.J.\ Bomans \inst{2}$^{\star,}$\thanks{Feodor-Lynen Fellow of the 
Alexander von Humboldt Foundation}
\and Y.-H.\ Chu \inst{2}$^{\star}$
\and M.D.\ Joner \inst{4}$^{\star}$
}

\offprints{K. Weis \\ email: kweis@ita.uni-heidelberg.de \\}

\institute{
Institut f\"ur Theoretische Astrophysik, Tiergartenstr. 15, D-69121 
Heidelberg, Germany
\and
University of Illinois, Department of Astronomy, 1002 W. Green Street, 
Urbana, IL 61801, USA
\and
Max-Planck-Institut f\"ur Radioastronomie, Auf dem H\"ugel 69, D-53121 Bonn, 
Germany 
\and
Brigham Young University, Dept. of Physics and Astronomy, 263 FB, Provo, 
UT 84602, USA
}

\date{Received 1 July 1996 /  Accepted 12 September 1996}

\maketitle

\begin{abstract}

HR Carinae is one of the few Luminous Blue Variables (LBVs) in the Galaxy.
It has a nebula that appears bipolar.  We have obtained imaging and 
high-dispersion, long-slit echelle data of the HR Car nebula, and confirmed 
that it is a bipolar nebula.  Its polar axis lies along the 
position angle of 125$\pm$5\degr; each lobe has, at a distance of 5\,kpc 
a diameter of $\sim0.65$\,pc and a line-of-sight expansion velocity 
of $75-150$\,${\rm km}\,{\rm s}^{-1}$.

Beside the expanding bipolar lobes, a number of [N\,{\sc ii}]-bright knots are 
detected.  These knots have lower expansion velocities than the lobes and
are detected only within the projected boundary of the lobes.  These knots 
are most likely nitrogen-enriched material ejected by HR Car.

On a larger scale, a funnel-shaped nebula is detected at 2$\farcm$5 northwest
of HR Car.  The axis of the funnel is roughly aligned with the polar axis
of the HR Car nebula, suggesting that HR Car may be responsible for the
ionization and shaping of this nebula.  Future observations of kinematics
and abundances are needed to determine the nature of this nebula.

We propose that the bipolar nebula of HR Car is a more evolved version of
the Homunculus Nebula around $\eta$ Car.  The recently developed
theory of wind-compressed disks may explain the higher density of the
equatorial plane and the formation of bipolar nebulae of LBVs.


\end{abstract}
\section{Introduction}
Luminous Blue Variables (LBVs) are often found in small circumstellar
nebulae, which contain the mass lost by these evolved supergiants 
(Nota et al.\ 1995).  With an initial mass $\ge$50\,M$_{\sun}$, 
these stars start as main-sequence O stars and evolve towards cooler 
temperatures at the end of hydrogen-core burning.  Following this track 
they enter the LBV phase when they approach the Humphreys-Davidson 
limit (e.g.\ Langer et al.\ 1994; Garc{\'\i}a-Segura et al.\ 1996).  The 
Humphreys-Davidson limit is an empirical upper boundary in the HR 
diagram (Humphreys \& Davidson 1979). Stars near this boundary are 
very unstable and show the highest mass-loss rates observed, 
several\,$\times10^{-4}$\,M$_{\sun}\,{\rm yr}^{-1}$. Losing more and more mass 
via continuous stellar winds and violent outbursts or eruptions, LBVs 
never reach the red supergiant phase but enter the Wolf-Rayet state 
after $\simeq 25000$\,yr in the LBV phase (Humphreys \& Davidson 1994).
The copious mass loss associated with the violent eruptions is 
responsible for the formation of a circumstellar nebula.  An excellent 
example is given by the LBV $\eta$ Carinae, around which the LBV nebula 
(LBVN) corresponds to eruptions mainly in $1840-1860$ (Polcaro and 
Viotti 1993; Viotti 1995).

The star HR Carinae (also known as HD 90177, SAO 238005, He\,3-407) is 
one of the few LBVs 
known in our Galaxy.  Its spectral type varies from B2\,{\sc i} to B9\,{\sc i};
strong Balmer, Fe\,{\sc ii} and [Fe\,{\sc ii}] emission lines are observed, 
with the Balmer 
and Fe\,{\sc ii} lines showing P Cygni profiles (Carlson \& Henize 1979; 
Hutsem\'ekers \& Van Drom 1991).  The distance to the star has been 
derived using two different methods which give consistent
results: kinematic distance $r_{\rm kin} = 5.4 \pm 0.4$\,kpc 
(Hutsem\'ekers \& Van Drom 1991), and reddening distance
$r_{\rm red} = 5 \pm 1$\,kpc (van Genderen et al.\ 1991).
Therefore, HR Car has a luminosity of 
$M_{\rm bol} = -9\fm 0 \pm 0\fm 5$, comparable to the other LBVs.  The 
circumstellar nebula around HR Car was not discovered until 1991, making 
it one of the newest member of LBVNs (Hutsem\'ekers \& Van Drom 1991).  
The origin and shape of the nebula around HR Car has been discussed by 
Hutsem\'ekers (1994).  A high-resolution image and a spectropolarimetric 
study have been presented by Clampin et al.\ (1995). 
 
We have obtained deep H$\alpha$ CCD images and high-dispersion, long-slit 
echelle observations of the nebula around HR Car.  The images are used to 
examine the structure of the circumstellar nebula and large-scale gaseous
environment of HR Car.  The echelle data are used to study the internal 
motion of the HR Car nebula.  In this paper  we report our analysis of the 
environment, kinematics, age, and evolution of the nebula around the
LBV HR Car.

\section{Observation and data reduction}

\subsection{Imaging}

Images of the field around HR Car have been taken with a CCD camera on 
the 0.9\,m telescope at the Cerro Tololo Inter-American Observatory in 
March 1996.  The seeing was between 1$\farcs$4 and 2$\arcsec$ during
the observations. An H$\alpha$ filter and Str\"omgren y filter were used.
The H$\alpha$ filter, having a central wavelength of 6563\,\AA\ and a 
filter width of 75\,\AA\,  included the [N{\sc ii}] lines at 
$\lambda$\,6548\,\AA \,  and $\lambda$\,6583\,\AA.
The Str\"omgren y filter was used to obtain continuum 
images to facilitate the continuum subtraction from the H$\alpha$ images.  
The $2048\times2048$ Tek2K3 CCD was used, and the pixel size was 
approximately 0$\farcs$4.  

To avoid saturation of the bright star HR Car itself, we secured
multiple short exposures.  A total of 10 frames in the y filter and 
60 frames in the H$\alpha$ filter were taken, each having 30\,s exposure.
To speed up the readout, only the central $1024\times1024$ pixels were 
read, hence the field of view was 6$\farcm75 \times 6\farcm$75.  
The combined H$\alpha$ image centered on HR Car is shown in Fig.\ 1.

A funnel-shaped nebulosity is detected to the northwest of HR Car in 
Fig.\ 1.  To investigate the spatial extent of this nebulosity, a second 
set of 5 H$\alpha$ and 5 y frames of 300\,s exposure each were taken of 
the adjacent field.  The combined H$\alpha$ image of the second field 
is displayed in Fig.\,2.  Note that the faint oval halo,
$\sim 2\farcm 5\times 2\farcm 1$, around HR Car in Fig.\ 1 is 
not present in Fig.\ 2, indicating that this halo is probably caused by 
the internal scattering inside the telescope optics rather than being 
real emission.

\begin{figure}
\epsfxsize=\hsize
\centerline{\epsffile{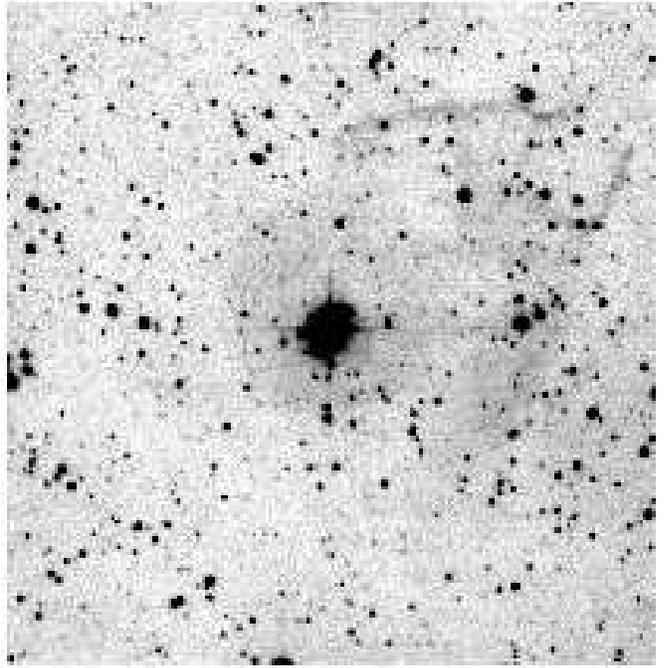}}
\caption{H$\alpha$ image centered on HR Car. The field of view is $6\farcm 75
\times 6\farcm 75$. North is up and east is to the left.}
\end{figure}

\begin{figure}
\epsfxsize=\hsize
\centerline{\epsffile{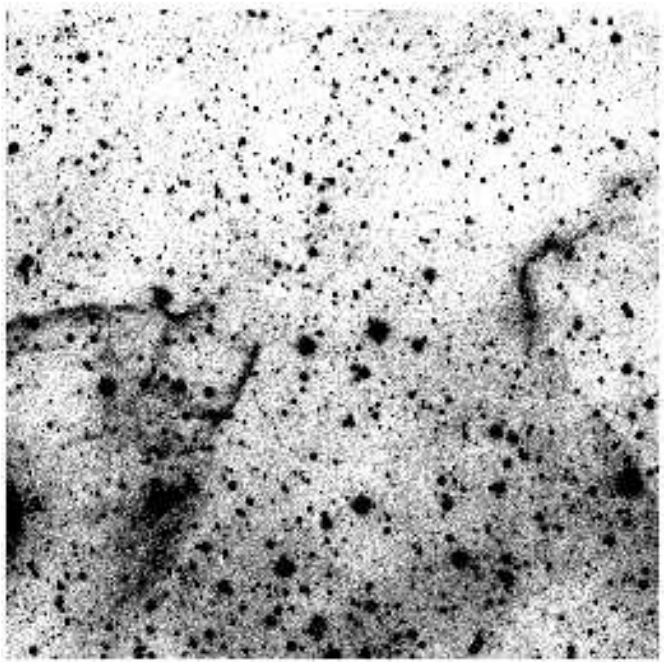}}
\caption{H$\alpha$ image centered at northwest of HR Car. The field of view 
is $6\farcm 75\times 6\farcm 75$. North is up and east is to the left.
HR Car is visible at the southeastern edge of the frame. }
\end{figure}

HR Car, the star itself, is so much brighter than the surrounding nebula
that it is necessary to subtract the continuum image from the H$\alpha$ 
image to obtain a better view of the nebula.  We have subtracted the 
scaled, combined y image from the combined H$\alpha$ image.  The scaling
factor is estimated using the neighboring stars.  HR Car is a strong 
H$\alpha$ emitter and hence cannot be removed completely in the 
continuum-subtracted H$\alpha$ image.  The central 
$60\arcsec\times 60\arcsec$ of the resultant H$\alpha$ image is 
displayed in Fig.\ 3.  The y filter and the H$\alpha$ filter have quite 
a large difference in their central wavelengths. Thus stars with extreme 
red or blue colors will be either under- or over-subtracted, and appear 
dark or white in Fig.\ 3.  The white arc to the north of HR Car is caused 
by an over-subtraction of the continuum caused by a low-intensity ghost in 
the y-filter images. 

\begin{figure}
\epsfxsize=\hsize
\centerline{\epsffile{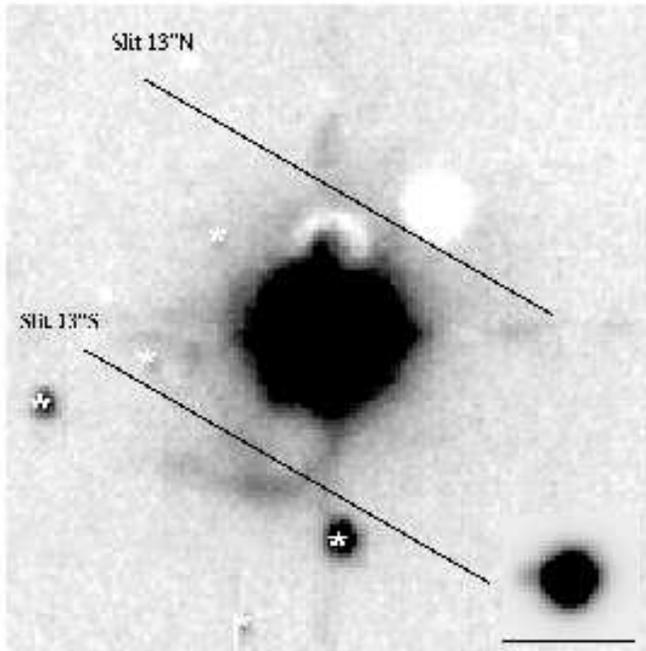}}
\caption{H$\alpha$ image of the nebula around HR Car. The field of view is 
60$\arcsec$ by 60$\arcsec$.  North is up and east is to the left.  The 
continuum is subtracted to our best effort. The over- and under-subtracted 
stars are the effect of their extreme colors.  HR Car has a strong stellar 
H$\alpha$ emission and hence remains bright in the continuum-subtracted image.
The inset at the lower right corner shows HR Car and its visual companion
at the same plate scale as the main image in this figure.}
\end{figure}

\subsection{Echelle spectroscopy}

We obtained high-dispersion spectroscopic observations of the nebula around 
HR Carinae with the echelle spectrograph on 
the 4\,m telescope at Cerro Tololo Inter-American Observatory in January 1996.
We used the long-slit mode, inserting a post-slit H$\alpha$ filter 
(6563/75\,\AA) and replacing the cross-disperser with a flat mirror.  A 
79\,l\,mm$^{-1}$
echelle grating was used.  The data were recorded with the long focus red
camera and the $2048 \times 2048$ Tek2K4 CCD.  The pixel size was 
0.08\,\AA\,pixel$^{-1}$ along the dispersion, and 0$\farcs$26\,pixel$^{-1}$ 
perpendicular to the dispersion.  The slit length was effectively limited
by vignetting to  $\sim4^\prime$.  Both H$\alpha$ 6563\,\AA\ and the [N\,{\sc 
ii}] 6548\,\AA, 
6583\,\AA\ lines were covered in the setup.  The slit-width was 250\,$\mu$m 
($\widehat{=} 1\farcs 64$) and the instrumental FWHM was about 
14\,km\,s$^{-1}$ at the 
H$\alpha$ line.  The seeing was $\sim 2\arcsec$ during the observations.
Thorium-Argon comparison lamp frames were taken for wavelength calibration
and geometric distortion correction.

Two slit positions were observed.  One was offset by 13$\arcsec$ to the 
north of HR Car, and the other 13$\arcsec$ to the south of HR Car.  For 
both positions the slit was rotated to a position angle of ${\rm PA} = 
60\degr$. The slit positions are marked in Fig.3.
The exposure time was 600\,s for the 13$\arcsec$\,S position and 
300\,s for the 13$\arcsec$\,N position.  The echelle images of the 
H$\alpha$+[N\,{\sc ii}] lines are presented in Fig.\ 4.  Unfortunately, the 
echelle observations were made before our discovery of the funnel-shaped 
nebulosity to the northeast of HR Car, so no echelle observations of this 
nebulosity are available.

\begin{figure*}
\epsfxsize=\hsize
\centerline{\epsffile{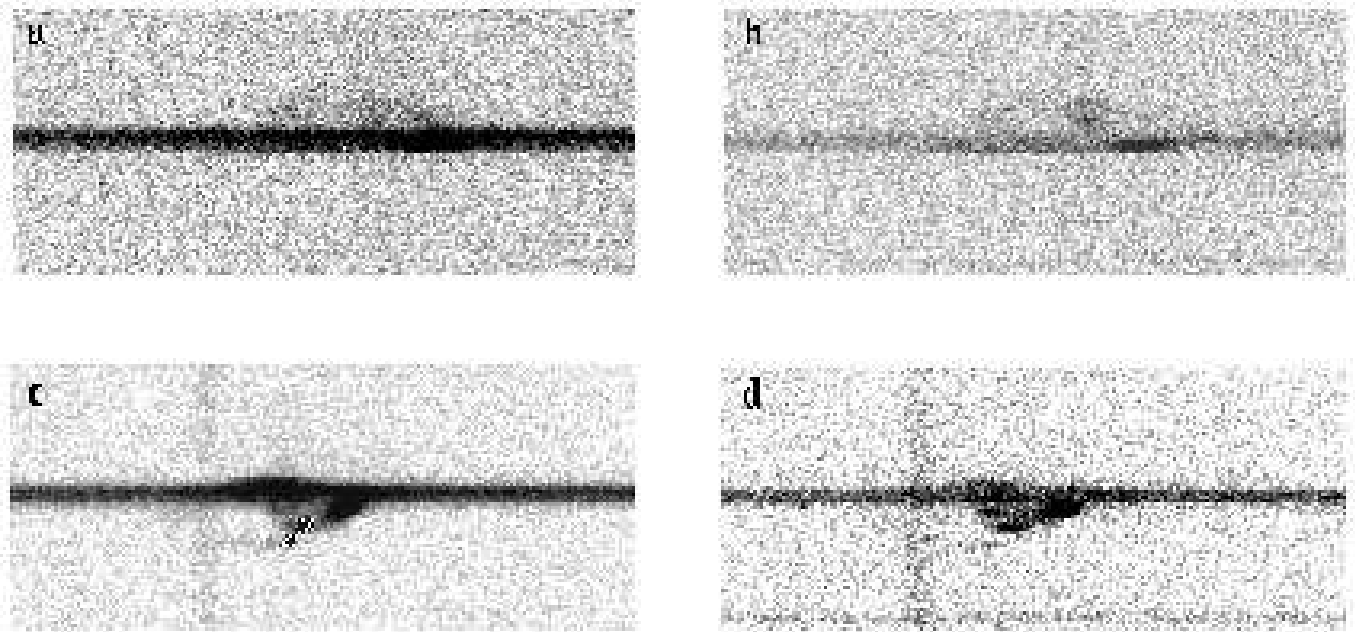}}
\caption{Echellograms of the HR Car nebula: (a) the H$\alpha$ line of 
the northwestern lobe, (b) the [N\,{\sc ii}]\,$\lambda$6583 line of the 
northwestern lobe, (c) the H$\alpha$ line of the southeastern lobe, and 
(d) the [N\,{\sc ii}]$\lambda$6583 line of the southeastern lobe. 
The horizontal axis is the position along the slit and covers 78\arcsec.  
Northeast is to the left and southwest is to the right.  The vertical 
axis is along the spectral dispersion and covers 13\,\AA, or
600\,km\,s$^{-1}$.  The wavelength (or velocity) increases upward.
The background H\,{\sc ii} component has $v_{\rm LSR} = -10.6$\,km\,s$^{-1}$.}
\end{figure*}

\section{The bipolar nebula of HR Car}

The nebula around HR Car has been classified as ``filamentary nebula'' 
by Nota et al.\ (1995) although a bipolar structure is suggested in 
Clampin et al.'s (1995) coronographic image, the best image of HR Car
available in literature.  In this coronographic image, the HR Car nebula 
appears elongated along the position angle $\sim 150 \degr$, with a 
size of approximately $37\arcsec\times 19\arcsec$.  The southeast lobe 
is brighter; it consists of a few nebular knots surrounded by a 
circular arc extending to $19\arcsec$ from the star.  The northwest lobe 
is fainter and less well-defined.

Our images of the HR Car nebula do not have as high a resolution as
the coronographic image, but our images are not occulted by a bar and
the stars are not saturated. Using the 
Str\"omgren y images, we identified stars near HR Car and marked 
the stars in Fig.\ 3 with asterisks.
We confirm the visual companion at 3$\farcs$5 east of HR Car
reported by Hutsem\'ekers \& Van Drom (1991).
This visual companion is about 6$^{\rm m}$ fainter than HR Car.
A y-band image of the HR Car and its companion is shown as an 
inset in Fig.\ 3.  

Some of the non-stellar emission features along the refraction spikes
are probably unreal, but the remaining diffuse emission features are real
and belong to the bipolar nebula of HR Car. The polar axis measured from 
the central
star and the southeast lobe is at ${\rm PA} = 125\pm$5\degr. 
The discrepancy of 25$\degr$ between this measurement and that 
measured from the coronographic
image is caused by the occulting bar which occulted the northern rim of
the southeast lobe in the coronographic image.  The size of the southeast
lobe measured from our images is $19\arcsec$ along the polar axis and 
$24\arcsec$ 
perpendicular to the polar axis.  The northwest lobe is faint; only the 
east rim of this lobe is easily seen in Fig.\ 3.

\section{Kinematics of the HR Car nebula} 

The long-slit echelle data were used to study the physical structure of
HR Car's nebula and to determine the radial velocities of the nebula
and its background H{\sc ii} region.
The brightest feature in the echelle images of the 
H$\alpha$ and [N\,{\sc ii}] lines is the extended background component at 
nearly 
constant velocity $v_{\rm LSR} = -10.6 \pm 2$\,km\,s$^{-1}$
The [N\,{\sc ii}]$\lambda$6583\,\AA/H$\alpha$ ratio of the background 
component is $0.30 - 0.35$, which is quite common for Galactic 
H\,{\sc ii} regions (Shaver et al.\ 1983).

The echelle images reveal two kinematic components in the HR Car nebula
(Fig.\ 4).  One component has a contiguous expansion structure and 
corresponds to the circular filaments, or the lobes.  The other component 
consists of knots expanding at slower velocities and corresponds to the 
knots encompassed by the circular filaments.  The expanding lobe component
has [N\,{\sc ii}]$\lambda\,$6583\,\AA/H$\alpha$ ratios of $0.30\pm0.05$, similar 
to those in the background H\,{\sc ii} component, while the knots have 
[N\,{\sc ii}]$\lambda\,$6583\,\AA/H$\alpha$ ratios
approaching 0.9, much higher than those in the lobes or the background 
H\,{\sc ii}. Hutsem\'ekers \& Van Drom (1991) obtained a 
[N\,{\sc ii}]$\lambda$\,6583\,\AA/H$\alpha$ ratio of 0.43, which may 
reflect a mixture of emission from lobe and knot components in their 
aperture.

The two kinematic components of the HR Car nebula are individually 
discussed below.

\subsection{The expanding lobes}

We have measured the radial velocities in H$\alpha$ and the 
[N\,{\sc ii}]$\lambda\,$6583\,\AA\ lines, and plotted them against positions 
along
the slit in Fig.\ 5.  The origin of the position-axis is where the slit 
intersects the polar axis (${\rm PA} = 150\degr$, as measured from the 
coronographic image). Offsets are positive to the southwest and negative 
to the northeast.  For reference, a dashed line is drawn at the velocity 
of the background emission, $v_{\rm LSR} = -10.6$\,km\,s$^{-1}$.

Comparisons between the continuum-subtracted H$\alpha$ image (Fig.\ 3)
and the echelle line images (Fig.\ 4) indicate that the filamentary
structure seen in the H$\alpha$ image is connected through a faint expanding
shell.  The circular filament extending to $19\arcsec$ southeast of HR Car 
corresponds to the rim of the expanding shell, or lobe.  The higher 
surface brightness of the circular filament is caused by the larger 
emitting depth along the line of sight.  The echelle images also 
reveal that the bipolar lobes of HR Car are larger than what we measured 
in our continuum-subtracted H$\alpha$ image (Fig.\ 3).  

The echelle slit positions were selected in assumption that the polar axis
was along ${\rm PA} = 150\degr$.  Fig.\ 5 shows that the largest expansion
velocity along the northern slit position occurs at about $5\arcsec$ 
southwest of the assumed polar axis, and that along the southern
position occurs at about $5\arcsec$ northeast of the assumed polar axis.
This position difference indicates that the position angle of the
polar axis should be smaller than ${\rm PA} = 150\degr$, supporting the
${\rm PA} = 125\degr \pm 5\degr$ measurement 
made with our H$\alpha$ images.

The velocity structure of the two lobes is typical for a bipolar 
expansion.  The northwestern lobe shows the largest receding 
velocity at $v_{\rm LSR} = +92 \pm 8$\,km\,s$^{-1}$ and the largest 
approaching velocity at $v_{\rm LSR} = -48 \pm 8$\,km\,s$^{-1}$, 
while the southeastern lobe 
shows the largest approaching velocity at $v_{\rm LSR} 
= -133\pm8$\,km\,s$^{-1}$ and the largest receding velocity at 
$v_{\rm LSR} = +30\pm8$\,km\,s$^{-1}$.  
If we assume point symmetry with respect to the star, these four extreme
velocities can be averaged algebraically to determine the systemic
velocity of the nebula, $\sim -14\pm5$\,km\,s$^{-1}$.
The interpretation or modeling of the velocity structure observed in the
lobes is complicated by three factors: (1) the lobes are not 
spherically symmetric, (2) the velocity vectors are not necessarily 
perpendicular to the lobe surface, and (3) the inclination angle of 
the polar axis is unknown.  Given the limited amount of echelle data, 
we can conclude that both lobes are expanding, but more observations
are needed for further modeling of the expansion.  Note that the 
approaching side of the southeast lobe may have different velocities in 
the H$\alpha$ and [N\,{\sc ii}] lines.  Unfortunately, our data do not have 
adequate S/N ratios to analyze this result further.

The systemic velocity of the nebula around HR Car
is close to the background H\,{\sc ii} velocity at $-10.6$\,km\,s$^{-1}$,
but very different from the LSR-velocity of $+9$\,km\,s$^{-1}$  reported by 
Hutsem\'ekers \& Van Drom (1991).  
Nevertheless our radial velocity is within the range of 
velocities measured by them for different ions in the stellar spectrum.
The kinematic distance implied by our systematic velocity of $v_{\rm LSR} = 
-14$\,km\,s$^{-1}$ is $2.3 \pm 0.5$\,kpc. Since peculiar velocities can affect
the kinematic distance, we will adopt the 5\,kpc. 

The background H{\sc ii} velocity is very similar to the velocity of 
the Carina 
nebula (Court\`es et al. 1966; Walborn \& Hesser 1975). This may be related
to the position of HR Car projected onto a faint, outer arc of the Carina 
nebula. The complex Na\,I interstellar absorption line profile as 
discussed by Hutsem\'ekers \& Van Drom (1991) with its components between 
+9 and $-99$\,km\,s$^{-1}$ can then be naturally explained by the motions 
inside the Carina H{\sc ii} complex (Walborn et al. 1984).

\begin{figure}
\epsfxsize=\hsize
\centerline{\epsffile{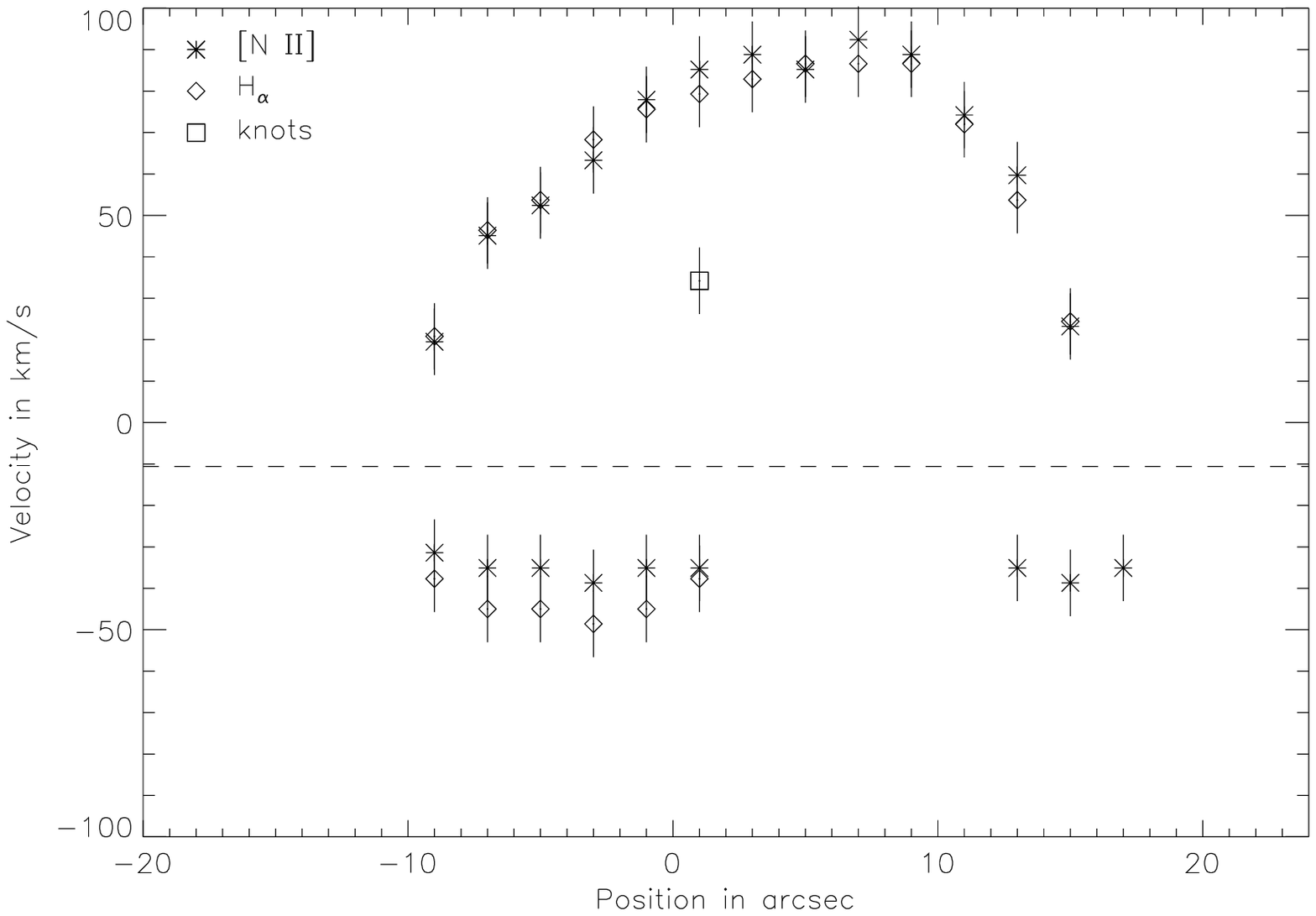}}
\epsfxsize=\hsize
\centerline{\epsffile{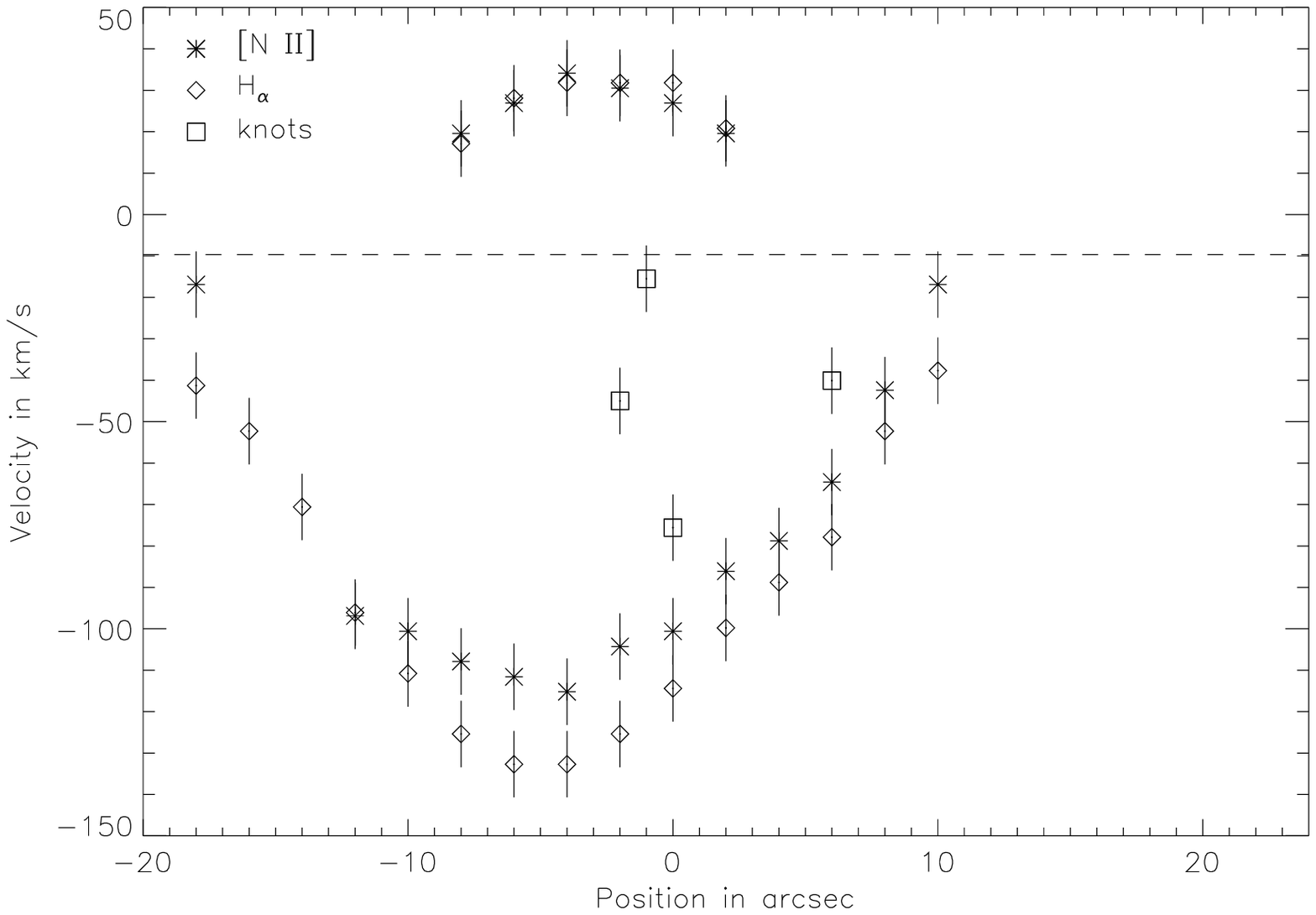}}
\caption{Velocity-position plots of the northern slit position (top)
and the southern slit position (bottom). Velocities are in LSR system.
The origin of the position-axis 
is at 9$\farcs$8 north and 5$\farcs$6 west of HR Car in the top plot,
and 9$\farcs$8 south and 5$\farcs$6 east of HR Car in the bottom plot. 
The slit is oriented along ${\rm PA} = 60\degr$.  The offsets are positive to 
the southwest and negative to the northeast.}
\end{figure}

\subsection{[N\,II]-bright knots}

Beside the main expansion structure, several [N\,{\sc ii}]-bright knots were 
detected in the echelle data.  We have identified 5 knots, 4 along the 
south slit and 1 along the north slit.  These knots are plotted as small 
squares in Fig.\ 5.  Their velocities (see Table 1) scatter between 
$-75$\,km\,s$^{-1}$ and $+34$\,km\,s$^{-1}$, always slower than the 
superimposed expanding lobe component.  These knots can be identified 
in the H$\alpha$ image in Fig.\ 3. All of these knots are inside or close 
to the bipolar main structure.  

Knots\,\#1$\,-\,$ \#4 are located in the southeastern lobe.  
Knots\,\#1 and \#2 
are projected near the center of the southeastern lobe and they expand 
slower than the lobe.  The different expansion velocities and the
different [N\,{\sc ii}]$\lambda$\,6583\,\AA/H$\alpha$ ratios indicate 
that these knots could not 
have the same origin as the expanding lobe.  It is not known whether
the knots are physically inside the lobes or are only projected within.
Knots\,\#3 and \#4 follow roughly the expansion of the lobe, but a few 
km\,s$^{-1}$ slower than the lobe.  The similarities in expansion 
velocity and projected position may be suggestive of physical interaction 
between the knots and the expanding lobe.

Knot\,\#5 is projected within the northwestern lobe. Like Knots\,\#1 and \#2,
Knot\,\#5 has an intermediate velocity that is slower than the expansion of
the lobe.  There is possibly another knot on the approaching side of the
northwestern lobe.  Kinematically it is not very different from the lobe, 
but the [N\,{\sc ii}]$\lambda$\,6583\,\AA/H$\alpha$ ratio is higher, 
$\sim 0.6$.

High [N\,{\sc ii}]$\lambda$\,6583\,\AA/H$\alpha$ ratios in 
LBV nebulae or knots are quite common. For example, the 
[N\,{\sc ii}]$\lambda$\,6583\,\AA/H$\alpha$
ratio for the S condensation of $\eta$ Car nebula maybe as high as a
$\simeq 7.5$ (Davidson et al. 1982). For AG Car, de Freitas Pacheco 
et al. (1992) found a value of 
[N\,{\sc ii}]$\lambda$\,6583\,\AA/H$\alpha = 0.48 \pm 0.05$ .

Similar [N\,{\sc ii}]-bright knots have been observed 
in ring nebulae around
Wolf-Rayet stars (e.g., RCW\,104, Goudis et al.\ 1988), but their origin
is unknown.  These knots are also reminiscent of the FLIERS seen in 
planetary nebulae, which have been suggested to be high-density, 
collimated material expelled recently from the nucleus (Balick et al.\ 
1994).  A high nitrogen abundance has been derived for [N\,{\sc ii}]-bright
Wolf-Rayet nebulae (M1-67, Esteban et al.\ 1991; NGC\,6888, Esteban \& 
V\'{\i}lchez 1992), as well as [N\,{\sc ii}]-bright knots in planetary nebulae
(NGC\,6571, Chu et al.\ 1991).  It is very likely that the [N\,{\sc ii}]-bright
knots in the HR Car nebula are overabundant in nitrogen.
The nitrogen enhancement fits well to the interpretation of HR Car being 
an LBV which lost processed material via outbursts. The overabundance of 
nitrogen follows naturally from the ongoing CNO cycle in the star, 
confirming LBVs to be evolved stars (Maeder 1983; Langer et al.\ 1994).

\begin{table}
\caption[]{Positions and velocities of the [N\,{\sc ii}]-bright knots.}
\begin{flushleft}
\begin{tabular}{clrrc}
\hline
knot & \multicolumn{2}{c}{lobe and} & velocity     & [N\,{\sc ii}]$
\lambda$\,6583\,\AA/H$\alpha$ \\
\#   & \multicolumn{2}{c}{position} & km\,s$^{-1}$ & \\
\hline
1    & southeast & $-2\arcsec$ &  $- 45.1$ & 0.88\\
2    & southeast & $-1\arcsec$ &  $- 15.5$ & 0.88\\
3    & southeast & $0\arcsec$ &  $- 75.6$ & $^*$  \\
4    & southeast & $6\arcsec$ &  $- 40.1$ & 0.83\\
5    & northwest & $1\arcsec$ &  $+ 34.2$ & 0.90\\
\hline
\end {tabular}
\\
\noindent $^*$ The H$\alpha$ line was affected by a cosmic ray hit, and
cannot be measured to give [N\,{\sc ii}]$\lambda$\,6583\,\AA/H$\alpha$ ratio.
\end{flushleft}
\end{table}

\section{Structure and environment }

We conclude that the nebula of HR Car is of bipolar structure and 
assume that the 
radial expansion velocity is no larger than the full velocity split 
and no smaller than 1/2 of the velocity split,
or 75\,km\,s$^{-1} < v_{\rm exp} < 150\,$km\,s$^{-1}$.
The northwestern and southeastern lobes have angular sizes of $26\arcsec$ and 
$28\arcsec$ along the slit, respectively.  For a distance of 5\,kpc, the 
linear sizes would be 0.63 and 0.67 pc.  Assuming no additional acceleration,
a lower limit on the dynamic age of the bipolar nebula is $4200 - 8400$\,yr.

The bipolar nebula of HR Car show intriguing similarities with the 
bipolar {\it Homunculus} nebula around the famous LBV $\eta$ Car.  
At a distance of 2.5\,kpc, each lobe of the Homunculus has a diameter 
of $\sim 0.1$\,pc.  This is a factor of 6 smaller than the lobes of
HR Car.  Therefore, it is very likely that the nebula around HR Car is 
an older version of the Homunculus around $\eta$ Car. 

The gaseous environment of HR Car can be seen in the H$\alpha$ images
in Figs.\ 1 and 2.  A large funnel-shaped nebula is visible to the
northwest of HR Car.  This nebula is very faint.  Only the brightest parts 
are barely visible on the ESO R plate in the Southern
Sky Atlas.  The axis of the funnel is roughly aligned with the 
the polar axis of the bipolar nebula around HR Car.  This morphology 
suggests that HR Car is responsible for the ionization and shaping
of this nebula.  

Is this funnel-shaped nebula ejected by HR Car or an interstellar bubble
blown by HR Car?  At a distance of 5\,kpc, the large shell structure would 
be located at 3 to 4.5\,pc from HR Car.  The ejecta of an LBV would 
need 26000 years to reach this distance from the star (assuming 
$v_{\rm exp} = 150$\,km\,s$^{-1}$).  This time interval is comparable
to the lifetime of the LBV phase.  Assuming $v_{\rm exp} = 240$\,km\,s$^{-1}$, 
a typical intermediate velocity of the shell from a LBV ejecta 
(Garc{\'\i}a-Segura et al.\ 1996), the filament at around 4\,pc would be 
reached within 16000\,yr. This would imply the possibility that 
this structure could be a remnant of an older ejecta from HR Car. 
If so, HR Car may be an already older LBV, 
which is consistent with its position in the HR diagram.  An older age is
implied by HR Car's being situated significantly beneath the 
Humphreys-Davidson limit (Humphreys \& Davidson 1994). 

Another possible scenario to connect this outer nebulosity with HR Car 
is to identify it as a wind-blown bubble. Fast stellar wind of massive stars 
in their main-sequence phase will sweep up ambient medium and form
``interstellar bubbles'' (Weaver et al.\ 1977).  The dynamic age of 
such a bubble is $\eta (R/v_{\rm exp}$), with $R$ being the bubble radius 
and $v_{\rm exp}$ its expansion velocity.  The parameter $\eta$ is 0.6 for 
an energy-conserving bubble in a homogeneous medium (Weaver et al.\ 1977) 
or 0.5 for a momentum-conserving bubble (Steigman et al.\ 1975).
Assuming a typical wind-blown bubble expansion velocity of 20\,km\,s$^{-1}$,
the dynamical age of this bubble (with the northern filaments being 
its border) would be between 1 and $ 2\times10^{5}$\,yr.  
With HR Car being already in the LBV phase a wind-blown bubble created in the 
main sequence state must be older than the $10^{5}$\,yr we find. If the 
northwestern nebulosity is connected to HR Car a wind-blown bubble scenario 
is not likely.

A net-like internal structure of this feature can very well be seen in Fig.2,
fitting to the scenario of an expanding shell structure.
We note that as seen in Fig. 2 this nebulosity shows an additional 
2$\arcmin$ e.g. 2.9\,pc long feature reaching towards the north-west. One 
may link this elongated cone to a blow out structure. To prove this and
explain the origin of the north-western filamentary nebula further 
spectroscopic and kinematic investigations are needed. 

Our echelle observations detect a background H$\alpha$ emission over
the entire slit length ($\sim 4^\prime$) at both slit positions.
The velocity
of this gas is $v_{\rm LSR} \sim -10.6$\,km\,s$^{-1}$ (see Fig.\ 4 and 
Fig.\ 5, the dashed line).  This background H\,{\sc ii} component has a 
velocity
FWHM of 41\,km\,s$^{-1}$ (after removing the instrumental broadening of
FWHM of 14\,km\,s$^{-1}$).  This velocity width, higher than the 
thermal broadening for a 10$^4$\,K gas, indicates gas motion; however,
there is no evidence for an expanding shell, since no line-split is
observed.  This could be caused by the presence of multiple unresolved
velocity components or a large turbulence. As discussed in 4.1 this may
be due to gas in the Carina H{\sc ii} Complex that is in line of sight to 
HR Car, or
indicates extended diffuse local emission around HR Car.

\section{Summary and Remarks}

Using imaging data and high-dispersion spectroscopic data, we have analyzed
the nebula around the LBV HR Car and confirmed its bipolar structure.
The two lobes have angular sizes of approximately 26$\arcsec$ and 28$\arcsec$  
and their expansion velocities amount to $75 - 150$\,km\,s$^{-1}$.  
Bipolar nebulae are actually not very common among the known LBVNs (Nota et 
al.\ 1995). 
The HR Car nebula will be the second bipolar LBVN ever reported, 
with the Homunculus Nebula of $\eta$ Car being the first. The formation 
of the bipolar nebula around HR Car is probably 
similar to that of the Homunculus nebula.  The Homunculus was ejected 
in the last century with an expansion velocity of $\simeq 700$\,km\,s$^{-1}$ 
(e.g.\,, Gaviola 1950; Meaburn et al. 1987; Meaburn et al. 1993; Damineli Neto 
et al. 1993). 
The LBV nebulae around HR Car and $\eta$ Car provide an opportunity to
probe the evolution of LBVNs. In such a scenario the formation of a 
LBVN starts with  
a nebula that looks more like the Homunculus (150\,yr old) 
and then evolves into a structure similar to
the bipolar nebula of HR Car ($4000 - 9000$\,yr old).
The formation and evolution of a LBVN may depend on the
strength of the eruption, initial expansion velocity, and the mass and 
density in the shell as well as in the surrounding medium.  Since so
many parameters are unknown, it is necessary to identify and use as many 
LBVNs as possible to complete our understanding of LBVN formation and
evolution.

Both LBVNs of HR Car and $\eta$ Car show bipolar morphology pinched by an 
equatorial plane.  This morphology may be supportive of the recent 
theory of wind compressed disks (Bjorkman and Cassinelli 1993; Mac Low et 
al.\ 1996).  In these models, the disk of higher density in the equatorial 
plan, formed by the wind and rotation of the star during the main sequence 
phase, is responsible for the mass ejection in preferred polar directions,
which lead to the formation of bipolar shaped LBVNs.  To verify this
formation mechanism, a better understanding of the kinematic structure 
of LBVN is needed.

\acknowledgements
{KW thanks Mordecai Mac Low, Guillermo Garc{\'\i}a-Segura, and Norbert Langer 
for stimulating discussions. We thank the referee, Damien Hutsem\'ekers, for 
his comments and suggestions. DJB acknowleges support from the Alexander 
von Humboldt Foundation through their Feodor Lynen Fellowship program.
MDJ thanks the Brigham Young University 
Department of Physics and Astronomy for continued support of his research
at CTIO.  We thank Peter W. A. Roming for work done during the initial
inspection of the optical imaging data, Richard Auer and Lisa Joner for 
editorial comments.}

\end{document}